# Plant survival and keystone pollinator species in stochastic coextinction models: role of intrinsic dependence on animal-pollination

Anna Traveset[1], Cristina Tur[1] and Víctor M. Eguíluz[2*]

[1]Institut Mediterrani d'Estudis Avançats IMEDEA (CSIC-UIB), Global Change Research Group, C/ Miquel Marqués 21, E07190-Esporles, Mallorca, Balearic Islands, Spain

[2]Instituto de Física Interdisciplinar y Sistemas Complejos IFISC (CSIC-UIB), E07122-Palma de Mallorca, Spain

*Corresponding author: victor@ifisc.uib-csic.es

**Abstract**

Coextinction models are useful to understand community robustness to species loss and resilience to disturbances. We simulated pollinator extinctions in pollination networks by using a hybrid model that combined a recently developed stochastic coextinction model (SCM) for plant extinctions and a topological model (TCM) for animal extinctions. Our model accounted for variation in interaction strengths and included empirical estimates of plant dependence on pollinators to set seeds. The stochastic nature of such model allowed us determining plant survival to single (and multiple) extinction events, and identifying which pollinators (keystone species) were more likely to trigger secondary extinctions. Consistently across three different pollinator removal sequences, plant robustness was lower than in a pure TCM, and plant survival was more determined by dependence on the mutualism than by interaction strength. As expected, highly connected and dependent plants were the most sensitive to pollinator loss and collapsed faster in extinction cascades. We predict that the relationship between dependence and plant connectivity is crucial to determine network robustness to interaction loss. Finally, we showed that honeybees and several beetles were keystone species in our communities. This information is of great value to foresee consequences of pollinator losses facing current global change and to identify target species for effective conservation.

Anthropogenic disturbances, such as habitat transformation, climate change or biological invasions, are at present main drivers of species loss and disruption of ecological interactions[1-3]. Ecological interactions among species are an important component of biodiversity because they provide relevant functions for populations, communities and ecosystems, such as pollination services[4-6]. For that reason, predicting responses of ecosystems and tolerance of species to disturbances is a key issue in ecology.

The extinction of a species inevitably causes the extinction of interactions, which in turn can trigger additional extinctions of species. Models with ecological interaction networks constitute a useful tool to simulate coextinction cascades, with the ultimate goal of understanding community robustness to species loss and resilience[7-10]. Topological coextinction models (TCMs) represented the first attempt to explore patterns of extinction in plant-pollinator networks[7]. Nevertheless, these models are mainly based on static network structure and have several important constraints. For instance, they assume that a species can only become extinct when all its interacting partners are lost; however, the primary loss of a pollinator species in real plant-pollinator networks may cause the coextinction of a plant, leading in turn to the coextinction of other pollinators that strongly depended on that plant, and even to the indirect coextinction of other plants which rely on those pollinators. TCMs also neglect interaction strength heterogeneity, assuming that after the loss of a species, all its partners are equally likely to coextinction; nevertheless, pollination interactions differ in terms of quantity and quality effects[11-13] and thus the coextinction probability of a species partner varies depending on the ecological effect of the lost interactions, *i.e.*, on its interaction strength[13,14]. Still, TCMs ignore that species vary greatly in their degree of functional dependence on mutualistic partners. For instance, in plant-pollinator networks, the plant breeding system is a key determinant of its dependence on pollinators to produce seeds and may modulate plant vulnerability to pollinator loss[15]. This dependence varies from high in plants with obligated cross-pollination to low in plants with autonomous self-pollination[16]. Thus, it seems crucial to take into account such dependence when modelling species vulnerability to coextinction.

Recently, a simple stochastic coextinction model (SCM) has been developed[17] that simulates species extinction cascades in mutualistic networks accounting for variation in interaction strengths and species dependence on the mutualism, thus relaxing the assumption that coextinctions require the loss of all partners. Simulations assumed three categories of dependence on the mutualism (low, intermediate or high)[17] due to the lack of empirical quantitative dependence estimates. More complex extinction cascades were obtained with this SCM than with TCMs and, unexpectedly, more coextinctions were found in highly connected networks.

In this study, we simulated pollinator extinctions in pollination networks and modelled coextinctions by using a hybrid model that combined the SCM[17] for plant extinctions and the TCM for animal extinctions. We used, for the first time, empirically-measured values of plant dependence on pollinators (IPD, hereafter; see methods to know how it was calculated) from two plant communities to estimate plant survival probability and robustness under three different extinction scenarios (pollinator removal sequences). We then compared our findings with those using TCMs. Moreover, we wanted to assess the relative importance of incorporating estimates of plant dependence on pollinators and heterogeneity in interaction strengths when simulating species extinctions. Still, we aimed at identifying the plant species that are more sensitive to pollinator extinctions and which pollinator losses cause the highest number of plant coextinctions.

**Results**

Plant robustness estimated with our hybrid model was significantly lower than with the TCM (ANOVA $F_{1,72} = 58.03$, $p < 0.001$; Fig. 1) in both study communities. This result was consistent regardless of the pollinator extinction scenario simulated (Supplementary Fig. S2). Specifically, the discrepancy in robustness between the two types of models was larger for highly connected and highly dependent plants on pollinators (Supplementary Fig. S3). TCM tended to overestimate robustness for generalist plant species, but to underestimate robustness for plants with a low dependency on animal pollinators.

When comparing the different model variants of SCM under the random extinction scenario, we found that models not incorporating plant dependence on pollinators (models D and DH; see methods) led to greater extinction cascades and smaller fractions of surviving species than models considering plant dependencies (models F and H; see methods) (Fig. 2). Consistently in the two communities, when half of the pollinators were extinct, the fraction of surviving plants was 40% in the D and DH models but c. 60% in the F and H models (Fig. 2b,d) because plants with some levels of self- or wind-pollination were able to tolerate pollinator losses. On average, 29% and 14% of plants, in the SB and the PM plant communities, respectively, survived after the extinction of all pollinators (models F and H). The lower fraction of surviving plants in PM compared to SB is probably a result of the higher overall IPD in PM (mean ± SD: 0.71 ± 0.24) than in SB (0.59 ± 0.38), although differences in average plant robustness between communities were not statistically significant ($F_{1,72} = 2.47$, $P = 0.12$). Species declines after pollinator extinctions obtained with model H, which considered dependence of plants on pollinators but simulated that all interaction strengths were homogeneous, were almost the same as those obtained with model F (Fig. 2). In addition, models D and DH, which also differed only in interaction strength heterogeneity, yielded similar results (Fig. 2). Overall, discrepancies between models show that plant dependence on pollinators is more important than interaction strength as determinants of plant species survival. Indeed, this result is supported with an analytical calculation for the random scenario. The survival probability of plant *i* after the removal of pollinator *j* is given by $P_{ij} = 1 - IPD_i \, d_{ij}$ (see Methods). If the pollinator to be extinct is selected at random then the expected survival probability of plant *i* will be obtained by averaging $P_{ij}$, over realizations of pollinator *extinctions j*, that is, $P_i = \langle P_{ij} \rangle = 1 - \langle IPD_i d_{ij} \rangle = 1 - IPD_i/N_P$, where $\langle \rangle$ is the average over realizations, and $N_P$ is the number of pollinators. For subsequent pollinator extinctions, a similar calculation leads to the result that the expected plant survival only depends on its corresponding dependence on pollinators and the specific connectivity pattern of the network (see Supplementary Information and Supplementary Fig. S1).

For each simulated extinction scenario (random, generalist and specialist), as pollinator extinctions go on, a lower survival probability is observed for plants, but survival probability collapses faster if plants depend more on pollinators (Fig. 3). Plants that do not rely on pollinators for reproduction (IPD = 0) were obviously insensitive to their loss and, thus, persisted in the network even after all pollinators had disappeared. On the other extreme, plants totally dependent on pollinators (IPD = 1) never persisted regardless of the pollinator extinction scenario simulated. In both communities, an important proportion (55.5% - 63.63%) of plant species was at high extinction risk, reaching survival probabilities below 0.1 during the different extinction cascades simulated. The effect of pollinator loss on plant survival was stronger under the 'generalist' simulation extinction scenario (sequential pollinator removal from the most to the least generalist) than in the 'specialist' scenario, whereas the effect was intermediate when extinctions occur randomly (Fig. 3). Plant survival probabilities decayed very fast under the 'generalist' scenario: after 5% of pollinator extinction events, plant survival probability was reduced by half for ca. 18% of plants in both communities, thus leading to a rapid species richness decline in the communities (Supplementary Fig. S4).

Plant survival probability was strongly negatively correlated with plant dependence ($r_s$ = -0.97, $P < 0.001$). Thus, plants highly dependent on pollinators became extinct before those depending little on them. Significant but lower negative correlations were also found between plant survival probabilities and plant centralities within the networks (degree: $r_s = -0.50$, $P < 0.05$; eigenvector centrality: $r_s = -0.41$, $P < 0.05$) suggesting a larger vulnerability to coextinction for plants with many generalist pollinators. Evenness of interactions, by contrast, was not associated to plant survival probabilities ($r_s = -0.20$, $P = 0.23$).

Regarding pollinators' traits, species leading to more coextinctions after their primary extinction were the most generalized pollinators, *i.e.*, those with more interactions with plants (Fig. 4; $r_s = -0.81$, $P < 0.001$). The fraction of surviving plants after pollinator extinction was also negatively correlated with pollinator species strength ($r_s = -0.88$, $P < 0.001$), evenness ($r_s = -0.63$, $P < 0.001$) and eigenvector centrality ($r_s = -0.61$, $P < 0.001$). In SB community, the pollinators causing the highest number of plant secondary extinctions were two beetles (*Meligethes* sp. and *Spermophagus* sp.) and the honeybee (*Apis mellifera*). In PM, on the other hand, the three keystone pollinator species were the beetle *Oedemera flavipes*, a small braconid wasp (*Chelonus* sp.)*,* and the honeybee. The honeybee was, thus, identified as a keystone species in both study communities leading to the coextinction of one plant species when it disappears from the communities (1.35 and 0.88 plant species on average in SB and PM, respectively). The loss of species from other functional groups, such as flies, bee flies or butterflies, had a minimum impact on either community (Fig. 5).

**Discussion**

Our study demonstrates that a key step to reach a more comprehensive understanding of species extinction risk in pollination mutualistic networks consists in incorporating real estimates of species dependence on partners into coextinction simulation models. We confirmed the finding by Ref. [17] that stochastic coextinction models (SCMs) predict lower network and species robustness than topological models (TCMs). This is because complex extinction cascades can occur when relaxing the assumption that coextinctions require the loss of all partners[17]. Nevertheless, extinction cascades under our hybrid model are constrained to be less complex than those under the SCM of Ref. 17 because pollinators can only go secondarily extinct when their last remaining partner disappears. Therefore, incorporating real estimates of pollinators' dependence on plants might even lead to a lower robustness than the observed here. We further found that discrepancy in robustness when comparing TCMs and SCMs was larger for generalized and highly-dependent species, which corroborates that extinction cascades are more likely under high dependencies and high connectance[17]. The fact that highly connected communities of mutualists are more sensitive to secondary extinctions than less connected communities challenges the old idea of a positive relationship between connectance and robustness to species extinctions in ecological networks[18-20] and at the same time calls for the search of mechanisms that allow species to persist in the community. Previous comparisons of simulations incorporating structural dynamics with those of static topologies in food webs showed also that species richness and connectance were uncorrelated with robustness[21]. In addition, our results can also be put in context of the idea of nestedness promoting stability in mutualistic networks. According to Ref. [22], the species contributing most to nestedness are those also contributing most to the persistence of the community, being in turn the most vulnerable to extinction. Given that the species contributing most to nestedness tend to be highly connected species, and considering our findings that these have a lower survival probability, we predict that species contributing most to nestedness will be those less likely to survive in a cascade of co-extinctions.

One interesting finding of our model was that plant persistence in the networks showed to be more influenced by their dependence on pollinators than by the heterogeneity in interaction strengths. At least this is true when using interaction frequency as a measure of interaction strength in networks. In general, interaction frequency is considered a good proxy of the total effect of animal mutualists on plants[14]. However, we must note that the heterogeneity of another measure of interaction strength representing the pollination qualitative component (*e.g.*, frequency of pollen deposition) rather than the quantitative one might be in some cases more relevant for determining plant survival. Moreover, note also that we averaged the results of the sequence of pollinator extinctions and this decreased the impact on plant survival; if the most important pollinator was always removed first, the impact would have been certainly greater. Models that neglect dependence represent the worst-case extinction scenario because they cannot simulate the persistence of plants that can set seeds without the help of pollinators. Given that natural communities, in general, vary in plant species dependence on the mutualism[15,23], it thus seems crucial to estimate and include such variable in models when assessing the relative robustness of different mutualistic communities to species extinctions.

Due to the increasing number of anthropogenic pressures on species, it is a major challenge to identify which life history, morphological or functional traits of species in ecological networks are associated to likelihood of extinction after partner loss. Our results show that, at least for plants in pollination networks, dependence on pollinators to produce seeds is a strong correlate of plant extinction probability and determines the order at which species go extinct. Plant species with a strong dependence, such as dioecious or self-incompatible species, were more prone to coextinction after pollinator losses and disappeared first from the community. On the other hand, low-dependent plant species, such as species with autogamous pollination, facultative selfing or wind-pollination, were less vulnerable and persisted longer in communities. Moreover, other plant species traits not studied here could contribute to the vulnerability or resilience of species to coextinction[24]. For instance, some plant species have alternative mechanisms to ensure reproduction without pollinators, such as vegetative propagation, which may also reduce their risk of extinction[16].

However, plant coextinction likelihood is not exclusively associated to plant reproductive dependence on pollinators, but also depends on their connectivity within the network. Consistently in the two communities, the most generalist plants tended to have lower survival probabilities. This is because plants with a high degree or centrality 'play the extinction lottery more times' when connected to more pollinators. Thus, generalized plant species that are self-incompatible and not wind-pollinated would be the most susceptible to pollinator losses and the first to disappear (*e.g.*, *Lotus corniculatus, Cistus salviifolius* in our communities). Interestingly, this suggests that the correlation between species dependence and connectivity in networks may be crucial to determine the robustness of entire communities to interaction loss. Communities can vary in the association between specialization level and dependence[23]. Although plants highly reliant on pollinators might be expected to have more interactions than less dependent species - to ensure their reproduction and buffer from fluctuations in pollinator populations-, the whole specialist-generalist gradient is found in both self-compatible and incompatible plants[15,25]. Communities where the most dependent species are also highly connected species (hubs), occupying central positions in the network, would be particularly susceptible to rapid coextinction cascades.

The SCM used here only incorporates real dependence estimates for plants, but ideally SCMs should add estimates of dependence also for pollinators. These estimates could be obtained, for instance, by measuring the contribution of pollen to the pollinators' offspring[26]. There are pollinators that strongly depend on floral resources for survival (obligate flower-visitors) because they solely feed on pollen and nectar, whereas others are less dependent (facultative flower-visitors) as they use pollen and nectar only as a diet complement[27]. Moreover, there are several ecological and life-history traits of pollinators, such as nesting, sociality or body size, which likely influence their sensitivity to disturbances[28]. Including this type of information in future models would certainly allow making better predictions on species coextinctions in pollination networks. Indeed, this framework could be extended to other kind of networks, such as seed-dispersal, host-parasitoid or plant-plant networks.

In addition, the identification of pollinator species that are more likely to trigger secondary extinctions can help targeting conservation efforts on keystone species[29]. Consistently in the two communities, pollinators leading to more coextinctions after their primary extinction were generalized species, occupying central positions in the network, showing a greater species strength, as well as a high interaction evenness. The loss of generalist pollinator species with high strength is likely to cause more coextinctions because there are more and stronger pathways available for the propagation of direct and indirect effects. A recent study actually found that deletion sequences which first removed species with the largest strength led to lower food web robustness than the simple removal of hubs[30]. Interestingly, the managed honeybee was identified as a keystone species, playing an important role for plant persistence in both networks studied, despite it represented a relatively small fraction of total pollinator visitation (ca. 11% and 18% in SB and PM, respectively). Its potential extinction or decrease in activity patterns due to the combined effect of pathogens, predators, pesticides, lack of floral resources, etc.[31] might cause secondary extinctions in pollination networks. The honeybee is widely known to be an invasive flower visitor, with high species strength, which is associated to its highly plastic behaviour, its efficient search and exploitation of resources, and its indiscriminate foraging[32]. The extinction of species from other functional groups showed a lower effect on plant survival, except for several species of beetles, which were also keystone species due to their ubiquity on flowers (*e.g.*, Ref. [6]).

Unfortunately, if keystone species are in turn more vulnerable to coextinction due to their higher connectance, rapid coextinction cascades could take place in communities. Previous studies[22] found that strong contributors to overall community persistence are in fact the nodes most vulnerable to extinction. However, experimental manipulative studies are needed[33,34] to verify theoretical predictions made by network coextinction models on species extinction risk and community robustness. Incorporating dependence on the mutualism and exploring the relationship between connectivity and dependence arises as a key point for future studies assessing community stability.

**Methods**

**Study sites and sampling of plant-pollinator interactions**

Plant-pollinator interactions were sampled in two different communities of Mallorca (Balearic Islands): (i) a dune marshland located at sea level in the northeast of the island (Son Bosc; SB hereafter) and (ii) a high mountain shrub located at ca. 1,100 m above sea level (Sa Coma de n'Arbona in Puig Major; PM hereafter). During two consecutive flowering seasons (years 2009 and 2010, from April to July at SB and from May to August at PM) time-fixed flower visitation observations (3 min in SB and 5 min in PM) were conducted on randomly selected flowering plants. Observations were performed between 10:00 am–5:00 pm on sunny and non-windy days. We recorded the identity of insect flower-visitors that contacted flower reproductive parts (pollinators, hereafter) and the number of flower visits made by each pollinator species. Pollinators unidentified in the field were captured for further identification by specialized taxonomists and were classified into 11 pollinator functional groups: large bees (≥10 mm long), small bees (<10 mm), honeybees, beetles, flies, hoverflies, bee flies, wasps, butterflies, ants, and others. More details about sampling and pollinators observed are in Ref. [23].

Flower-visitation data from both years were pooled to construct quantitative plant-pollinator networks for each study site only including species for which reproductive dependence on pollinators was studied (see next section). Interaction strength in such networks was the total number of visits per flower per hour. These networks were later used to simulate coextinction cascades.

**Measuring plant reproductive dependence on pollinators**

For ca. 30% of flowering plant species in each study community we estimated reproductive dependence on pollinators as the difference in seed set with and without pollinators. The study species (SB = 27, PM = 11) represented 42% and 35% of all plant families present in both communities, respectively, and included both specialist and generalist plants (from one or a few pollinators to many). When plants had flowers at bud stage, we randomly selected 3-4 plant individuals per species. On each plant we counted and marked flowering branches or pedicels to conduct two treatments: (i) open pollination (OP), *i.e.*, flowers left to be naturally pollinated, and (ii) pollinator exclusion (PE), *i.e.*, flowers covered with fine mesh bags preventing insect visitation but allowing pollen grains to go through, to assess the role of self-pollination and the potential effect of wind. The number of flower units marked per plant and treatment varied depending on species floral display. Plants were weekly monitored, and fruits set in each treatment were collected when nearly or fully ripe. Fruits were dissected in the laboratory and viable seeds were counted. Mean seed set (*SS*) for each treatment was calculated as the total number of seeds produced per marked flower unit.

The degree of plant dependence on insect pollinators (*IPD*) was estimated as the fraction of seed set attributable to pollinator interactions, *i.e.*, the open pollination seed set after excluding the contribution of self- and wind-pollination. Thus, for each plant species we calculated *IPD* as $IPD = \frac{SS_{op} - SS_{pe}}{SS_{op}} = 1 - \frac{SS_{pe}}{SS_{op}}$, being $SS_{op}$ and $SS_{pe}$ the mean seed set obtained in open pollination and in the pollinator exclusion treatment, respectively. Therefore, *IPD* ranges from 0 for plants that produce seeds independently of pollinators (*i.e.*, by selfing and/or wind-pollination) to 1 for plants fully relying on pollinators for seed production. For most species, *IPD* may be contingent upon the biotic and abiotic conditions of each particular site, and may also vary across seasons, although for other species, *IPD* can be considered as constant (*e.g.*, those that are self-incompatible and are not wind-pollinated have a consistent *IPD* = 1).

**The coextinction stochastic model**

We adapted the recently developed stochastic simulation model for coextinctions in mutualistic networks[17] that considers the intrinsic reproductive dependence of species on the mutualism and the relative frequency of each interaction between mutualistic partners (dependence sensu Ref. [35]).

Our model simulates species extinction cascades and begins with the removal of a pollinator species *j* in the network. After the primary extinction of a pollinator *j*, all their plant partners have a probability of survival given by $P_{ij} = 1 - IPD_i\, d_{ij}$, where $P_{ij}$ is the survival probability of plant species $i$ following the extinction of a partner pollinator species $j$; $IPD_i$ is the empirical estimate of reproductive dependence on insect pollinators of plant species $i$ (*i.e.*, to what extent pollinators contribute to $i$'s seed set); $d_{ij}$ is defined as the fraction of all pollinator interactions involving plant $i$ and pollinator $j$, and was estimated as the fraction of visits by $j$ to $i$ divided by the total visits of the surviving partners to $i$ and including $j$ itself (following Refs. [17,35]). Thus, plant coextinction is more likely (small $P_{ij}$) when plants strongly depend on insect pollination to set seeds and the pollinator partner lost is important in terms of interaction frequency. When secondary plant extinctions occur, they can in turn lead to new pollinator extinctions. Because empirical estimates of the reproductive dependence of pollinators on plant food resources (pollen, nectar) were not available in our study, the survival probabilities of pollinators are not calculated with the previous formula. Instead, in our model, a pollinator can only become extinct when it is left without food resources, *i.e.*, it has lost all interacting partners. The loss of pollinators that interacted with an extinct plant therefore cannot lead to more plant extinctions. Thus, the model we used may be viewed as a hybrid version between stochastic and topological models (stochastic for plants and topological for pollinators).

When no more coextinctions occur after each primary pollinator extinction, the number of plants and pollinators remaining in the network are calculated, and a new pollinator extinction is simulated again, and so on, until no pollinators are left in the network (see pollinator extinction simulation scenarios). Note that interactions that are lost after primary extinction are not recovered after each secondary extinction step. As the extinction cascade goes on, new survival probabilities for plants are estimated each time after an extinction event by recalculating relative interaction strengths $d_{ij}$. In other words, after each pollinator loss, the rest of pollinators of a given plant "are allowed" to compensate the number of visits of that lost pollinator by increasing their interaction strengths with the plant[36]. For example, species $i$ has interaction strength $d_{ij} = 1$ when $j$ is its last surviving partner, regardless of the initial value of $d_{ij}$. In this case, $P_{ij} = 0$ when the plant fully depends on pollinators, but $P_{ij} > 0$ when $IPD_i \neq 1$. Therefore, plant species $i$ may persist even if it has lost all of its partners, an important difference from the static topological models. However, our model does not consider re-wiring, *i.e.*, species cannot establish new interactions after the loss of their original mutualistic partners.

**Pollinator extinction simulation scenarios**

For each network, we simulated three different scenarios of pollinator species loss with our model: (1) *random* scenario, in which pollinator species were removed in a random order, (2) *generalist* scenario, in which pollinator species were sequentially removed from the most to the least connected pollinator, and (3) *specialist* scenario, in which pollinator species were sequentially removed from the least to the most connected. The random scenario represents a null scenario with which to compare, as done in other studies[7,20]. The generalist scenario is typically used to explore the "attack tolerance" of networks to the loss of hub nodes[20,37,38]; this scenario may occur, for instance, when super-generalist invasive species (which sometimes can be alien invasive species[32,39,40]) are removed in communities[41]. Lastly, the specialist scenario is expected when weakly linked species are at greatest risk of real-world extinction. For instance, specialist pollinator species have been found to be more vulnerable than generalist ones to land-use intensification and habitat fragmentation[42,43].

The simulations of the stochastic coextinction model were implemented in FORTRAN (v. f 95). A total of 10,000 sequences of pollinator removal with no replacement were performed for each scenario. Plant survival probabilities were obtained after each extinction event (pollinator species removal) and after the extinction of all pollinators under each simulated scenario. We also calculated robustness ($R_i$) for plant species $i$ as the area below the decaying curve of the survival probability of $i$ after each pollinator extinction (measured as the fraction of extinction events relative to the total number of pollinators). Plant robustness is a quantification of plant's tolerance to pollinator loss. The value is bounded between 0, when plant survival probability decreases abruptly after pollinator extinctions, and 1, when it decreases mildly, thus indicating fragility and robustness, respectively[44].

**Adding complexity to coextinction simulation models**

We were interested in determining how the addition of estimates of plant reproductive dependence on pollinators and quantitative interaction strengths in coextinction simulation models may affect conclusions drawn about species robustness to pollinator loss. Hence, we first compared plant robustness estimates obtained with our SCM to those obtained with TCM. Secondly, we assessed the relative importance of including dependence on pollinators and the heterogeneity of interaction strengths in our simulation coextinction model by comparing, for the random scenario, the results obtained with different variants of the SCM (F, D, H and DH). Model F corresponds to the full coextinction model, as explained above, which incorporates both empirical estimates of plant reproductive dependence on pollinators (IPD) and interaction strengths. In model D, we simulated a community in which all plant species depended entirely on animal pollination to set seeds, *i.e.*, we assigned all species in the network an $IPD_i = 1$, but kept observed interaction strengths. In model H, there was no heterogeneity in interaction strengths, *i.e.*, we assigned homogeneous interaction frequencies among pollinators of plant *i* but maintaining $IPD_i$ to the empirical values. Finally, model DH simulated that all plants depended entirely on pollinators ($IPD_i = 1$) and that interaction strength was homogenous across all pollinator species of plant *i*. The decay in species and plant richness with pollinator loss was compared among the four model variants.

**Sensitivity of plant species to pollinator loss and identifying keystone pollinators**

Plant survival probability after the extinction cascade simulations gives an idea of sensitivity of plant species to loss of pollinator partners. Extinction risk is likely to depend on species traits[24,45] and network characteristics[46]. In order to determine which traits were associated to the sensitivity of plant species to pollinator loss, we assessed the relationship between plant survival probabilities and: (a) plant dependence of pollinators, (b) plant degree (normalised by the total number of pollinators in each network), (c) interaction evenness, and (d) eigenvector centrality scores, which measure how central is a species in the network taking into account the centrality of partner species, so that plant species that tend to interact with more generalized pollinators will have larger scores[47,48]. Spearman rank correlations between variables were conducted pooling data for the two communities.

Finally, we identified pollinator keystone species in our communities, *i.e.*, pollinators whose loss would cause more coextinctions. In order to do that, we simulated with SCM (10,000 simulations) the selective extinction of each pollinator species in our networks and compared the number of surviving plant species after pollinator removal. We then assessed whether this fraction of surviving plants was associated to: (a) pollinator degree (normalised by the number of plants in each network), (b) interaction evenness, (c) eigenvector centrality, and (d) species strength[35]. Spearman rank correlations between these variables were conducted.

**Data Availability**

The datasets analysed during the current study as well as figure source data and scripts are available in the digital.csic repository [http://dx.doi.org/10.20350/digitalCSIC/8507].

**Acknowledgments**

We thank Rocío Castro-Urgal, Jaume Reus, Pep Mora, Joan G. Torrandell, Héctor Guerrero and Zeeba Khan for their work in the field during pollinator surveys and flower bagging experiments, and Jaume Reus, Waleska Vázquez and Clara Vignolo for helping in the laboratory counting seeds. We are also very grateful to the taxonomists that identified the collected insects: David Baldock, Xavier Canyelles, Leopoldo Castro, Andreas Werner Ebmer, Xavier Espadaler, David Gibbs, Gerard Le Goff, Jordi Ribes, Paolo Rosa and Erwin Scheuchl. This study was performed within project CGL2013-44386-P financed to AT by the Spanish Ministry of Economy and Competitiveness, and project SPASIMM (FIS2016-80067-P AEI/FEDER, UE) to VME by Agencia Estatal de Investigación (AEI, Spain) and Fondo Europeo de Desarrollo Regional (FEDER).

**Author contribution**

The three authors designed the study. VME developed the model and run all the simulations. CT and AT compiled the network data from our previous studies. CT performed the statistical analyses and drew the figures. AT led the writing, with substantial contributions from all authors.

**Competing financial interests.** Authors declare no competing financial interests.

**Materials & Correspondence:** Dr. V. M. Eguíluz

**FIGURES**

**Fig. 1. Plant robustness.** Mean and 95% confidence intervals of plant robustness estimated both with the topological coextinction model (TCM) and the stochastic coextinction model (SCM) in the two communities (SB: Son Bosc, PM: Puig Major). Data correspond to the random extinction scenario. In each community, confidence intervals do not overlap and plant robustness with the SCM was lower than with TCM.

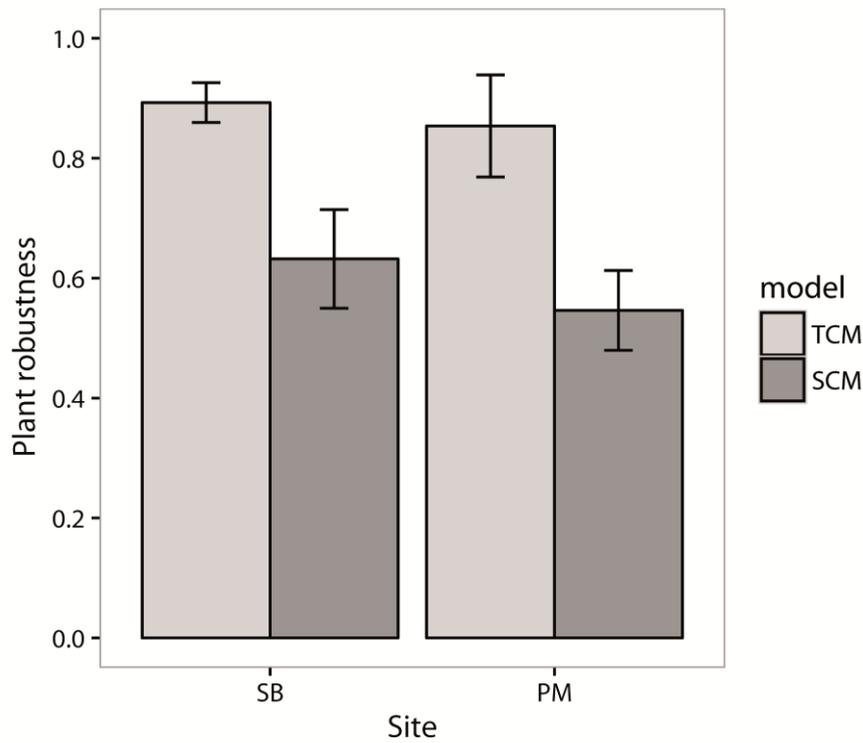

**Fig. 2. Surviving species as a function of extinction events.** Fraction of surviving species (a, c) and surviving plant species (b, d) resulting from pollinator extinctions under the random scenario in SB (above) and PM (below) communities. Model F simulates extinctions considering the variation in both IPD (plant dependence on pollinators; see methods) and interaction strength. In model D, all plants have an IPD=1, *i.e.*, depend entirely on animal pollination to set seeds. Model H simulates the case in which there is no heterogeneity in interaction strength. Finally, in model DH all plants have an IPD=1 and there is homogeneity in interaction strength. Differences between models that do not consider the real dependence of plants on pollinators (F *vs.* D, H *vs.* DH) are higher than between models that neglect interaction strength heterogeneity (F *vs.* H, D *vs.* DH). Note that the curves of the fraction of surviving species and plant species are on top of each other for Models F and H (maintaining the empirical IPD), and are also for Models H and DH (when plants depend entirely on pollinators). 95% confidence intervals are of size of the line width and are not shown for clarity.

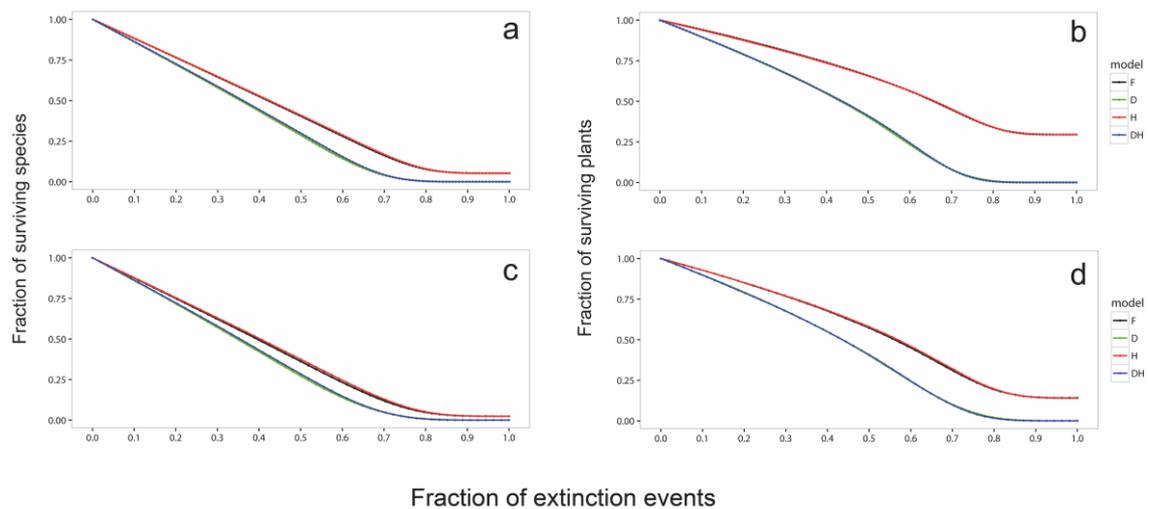

**Fig. 3. Dependence on pollinators and plant survival.** Heatmaps showing the variation in plant survival probability according to plant dependence on pollinators through the cascade of extinction events under the three simulated scenarios with our full model: random extinctions, extinctions from the most to the least generalist species, and extinctions from the least to the most generalist species. In each matrix plant species (rows) were ranked by decreasing dependence on pollinators (IPD). As the extinction cascade goes on plant survival probabilities decrease, but plants with low dependencies on pollinators to set seeds (IPD → 0) tend to be more persistent that plants with a high dependence (IPD → 1) on pollinators.

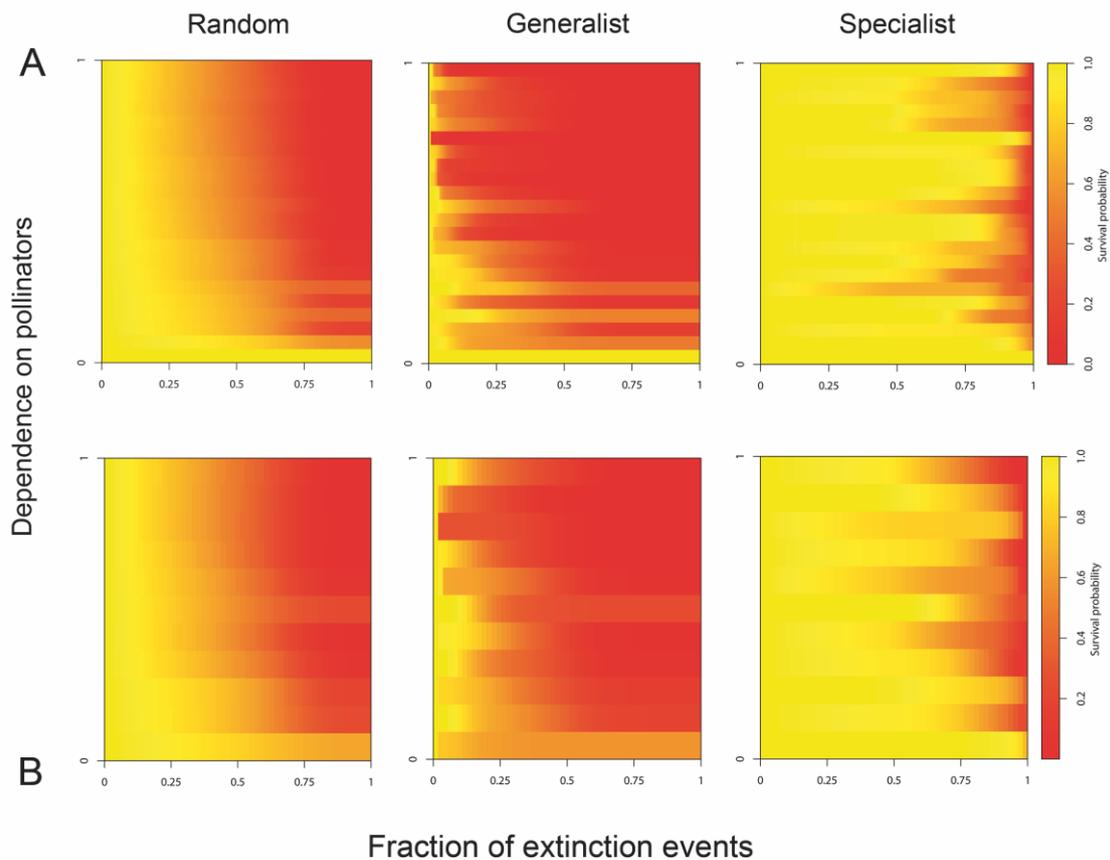

**Fig. 4. Pollinator degree.** Association between pollinator degree and the number of surviving plants after extinction of each pollinator in the two communities: (a) Son Bosc and (b) Puig Major. The loss of highly connected pollinator species caused the largest number of coextinctions in both networks.

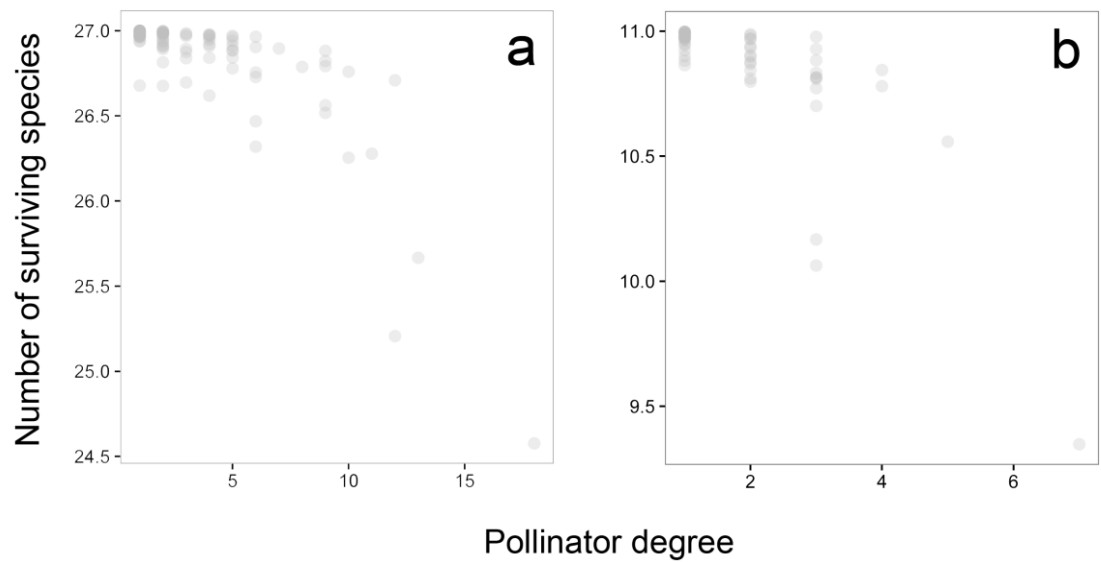

**Fig. 5. Functional groups influence on surviving plants.** Mean fraction of surviving plant species after the extinction of pollinators belonging to different functional groups. BF: bee flies, BU: butterflies, OT: others (mostly orthopterans and hemipterans), FL: flies, WA: wasps, LB: large bees, SB: small bees, HF: hoverflies, BE: beetles, AN: ants, HB: honeybee. The strength of each pollinator species is also depicted in a colour scale. The dashed lined represents the average effect of losing a single pollinator species. Data are shown for SB (a) and PM (b) communities. Note that the losses of species with highest strength are those causing the largest plant extinctions.

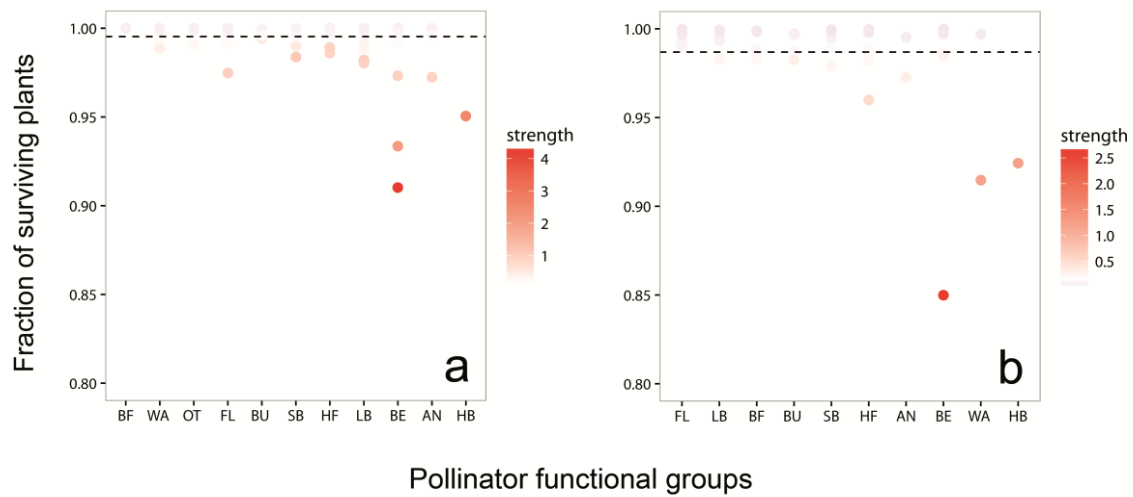

Supplementary Information

# Plant survival and keystone pollinator species in stochastic coextinction models: role of intrinsic dependence on animal-pollination

**Anna Traveset[1], Cristina Tur[1] and Víctor M. Eguíluz[2*]**

[1]Institut Mediterrani d'Estudis Avançats IMEDEA (CSIC-UIB), Global Change Research Group, C/ Miquel Marqués 21, E07190-Esporles, Mallorca, Balearic Islands, Spain

[2] Instituto de Física Interdisciplinar y Sistemas Complejos IFISC (CSIC-UIB), E07122-Palma de Mallorca, Spain

*Corresponding author: victor@ifisc.uib-csic.es

**Calculation of the survival probability of plants**

At each extinction event one pollinator is selected at random for extinction. The probability that plant *I* survives is given by

$$P_{ij} = 1 - IPD_i d_{ij}, \tag{S1}$$

where $IPD_i$ is the dependence on pollinators of plant *i*. Averaging over realizations leads to the expected surviving probability of plant *i*,

$$P_i(1) = \langle P_{ij} \rangle = 1 - IPD_i \langle d_{ij} \rangle = 1 - \frac{IPD_i}{N_P}, \tag{S2}$$

where $N_p$ is the number of pollinators. For the second extinction, we select a random pollinator *k* from the surviving set. Assuming that the first extinction has not led to tertiary extinctions, the probability of plant *i* to survive to 2 extinctions is given by

$$P_i(2) = \langle (1 - IPD_i d_{ij})(1 - IPD_i d'_{ik}) \rangle = \left(1 - \frac{IPD_i}{N_P}\right)\left(1 - \frac{IPD_i}{N_P - 1}\right). \tag{S3}$$

For *e* extinctions we can obtain

$$P_i(e) = \langle (1 - IPD_i d_{ij})(1 - IPD_i d'_{ik}) \rangle = \left(1 - \frac{IPD_i}{N_P}\right)\left(1 - \frac{IPD_i}{N_P - 1}\right) \ldots \left(1 - \frac{IPD_i}{N_P - e + 1}\right). \tag{S4}$$

For plants with $IPD_i \neq 1$, this can be expressed as

$$P_i(e) = \frac{\Gamma(N_P + 1 - IPD_i)}{\Gamma(N_P + 1 - IPD_i - e)} \frac{\Gamma(N_P + 1 - e)}{\Gamma(N_P + 1)}, \tag{S5}$$

where $\Gamma(.)$, is the Gamma function; while for $IPD_i = 1$,

$$P_i(e) = \frac{N_P - e}{N_P}. \tag{S6}$$

Supplementary Figure S1 shows the comparison between the numerical and the theoretical values. For the theoretical values the fraction of surviving plants is obtained using Equation (S4) for the empirical values of SB (top) and PM (bottom). The discrepancies can be explained by the ansatz of absence of pollinator extinctions after a plant extinction, which depends on the connectance of the plant-pollinator network.

**Supplementary Figure S1. Fraction of surviving plants as a function of the fraction of extinction events.** The numerical response (triangles) is compared with the theoretical value (squares) for the complete model (model F) and for the case where all the IPD values are set to 1 (model D), (a) SB site and (b) PM site. The theoretical curves are obtained averaging Equations (S5) and (S6) with the empirical IPD values.

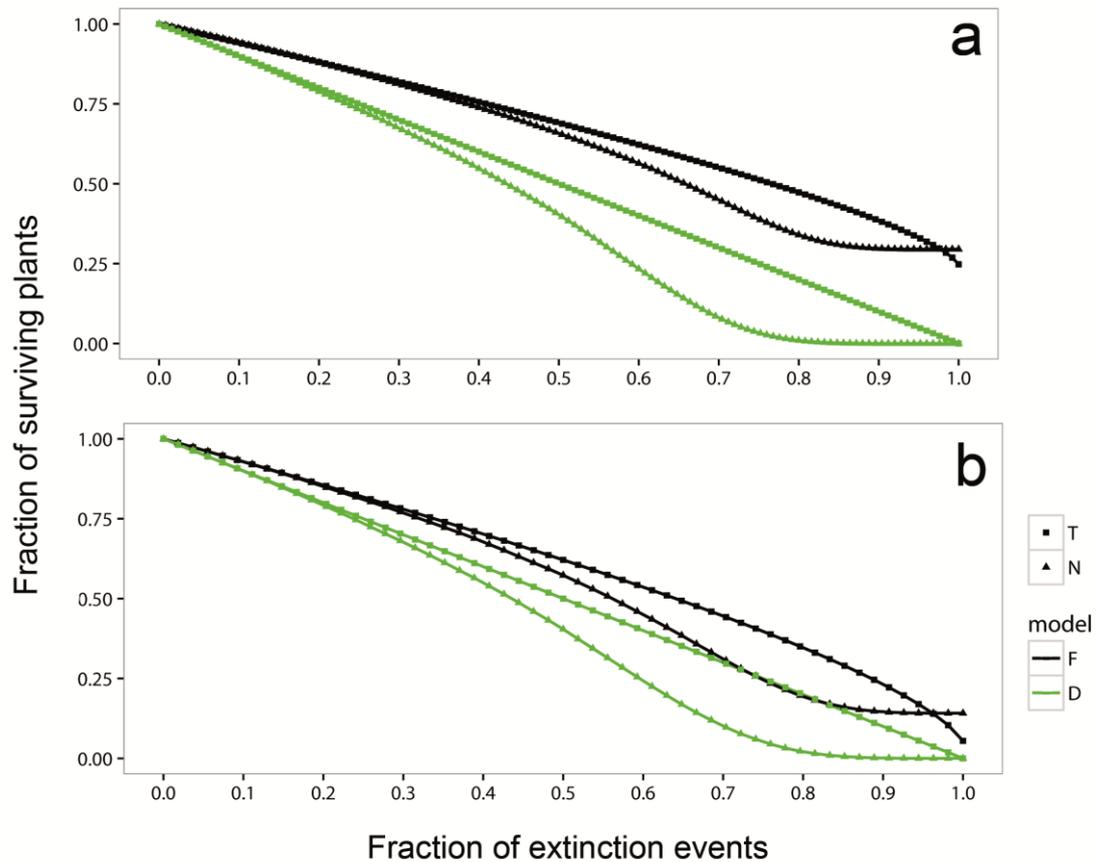

**Supplementary Figure S2. Plant robustness *vs.* extinction models.** Mean and 95% confidence intervals of plant robustness estimated both with the topological coextinction model (TCM) and the stochastic coextinction model (SCM) in the two communities (a: Son Bosc, b: Puig Major) under the three pollinator removal scenarios (R: random, G: generalist, S: specialist). Consistent significant differences were found between the two models in all scenarios.

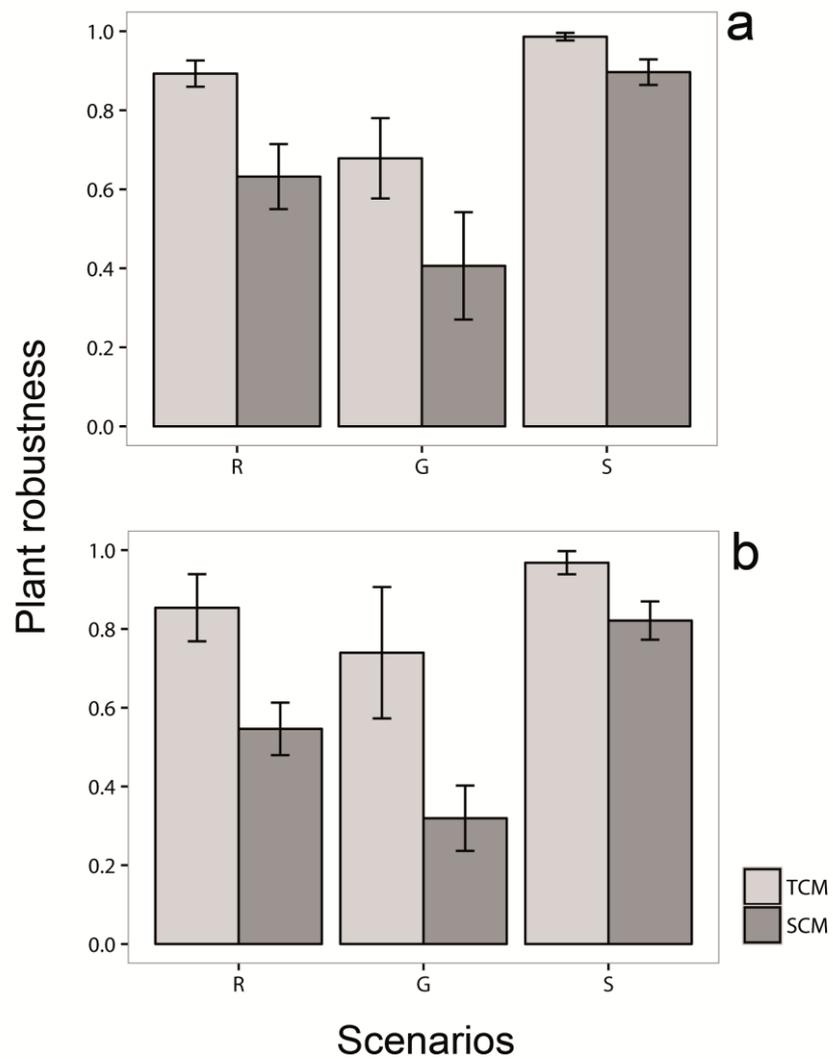

**Supplementary Figure S3. Plant robustness *vs.* degree and pollinator dependence.** Difference in plant robustness among the topological coextinction model (TCM) and the stochastic coextinction model (SCM) according to: (a) plant degree, and (b) plant dependence on insect pollinators (IPD). Each dot is a plant species and colours indicate the sampling community. For plants above the line, robustness was higher when estimated with TCM than with SCM. TCM tends to underestimate robustness for plants with few interactions (small degree) and plants with a low dependence on pollinators to produce seeds.

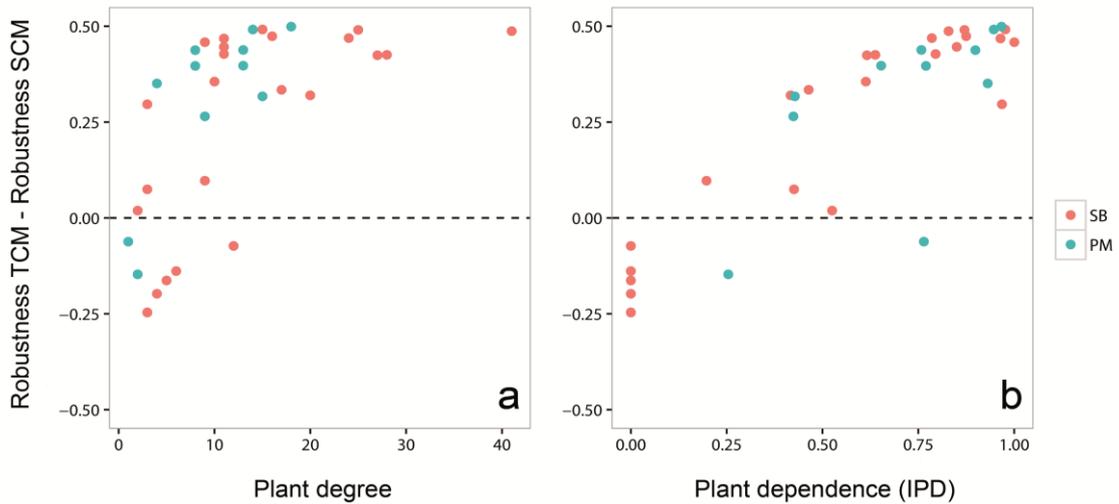

**Supplementary Figure S4. Surviving species depending on extinction scenario.** Fraction of surviving species (a,c) and surviving plant species (b,d) resulting from pollinator extinctions with the full model under the three extinction scenarios in SB (above) and PM (below) communities. R: random scenario, extinctions occur randomly across pollinators; G: generalist scenario, extinctions go sequentially from the most to the least linked species; S: specialist scenario, extinctions go sequentially from the least to the most linked species.

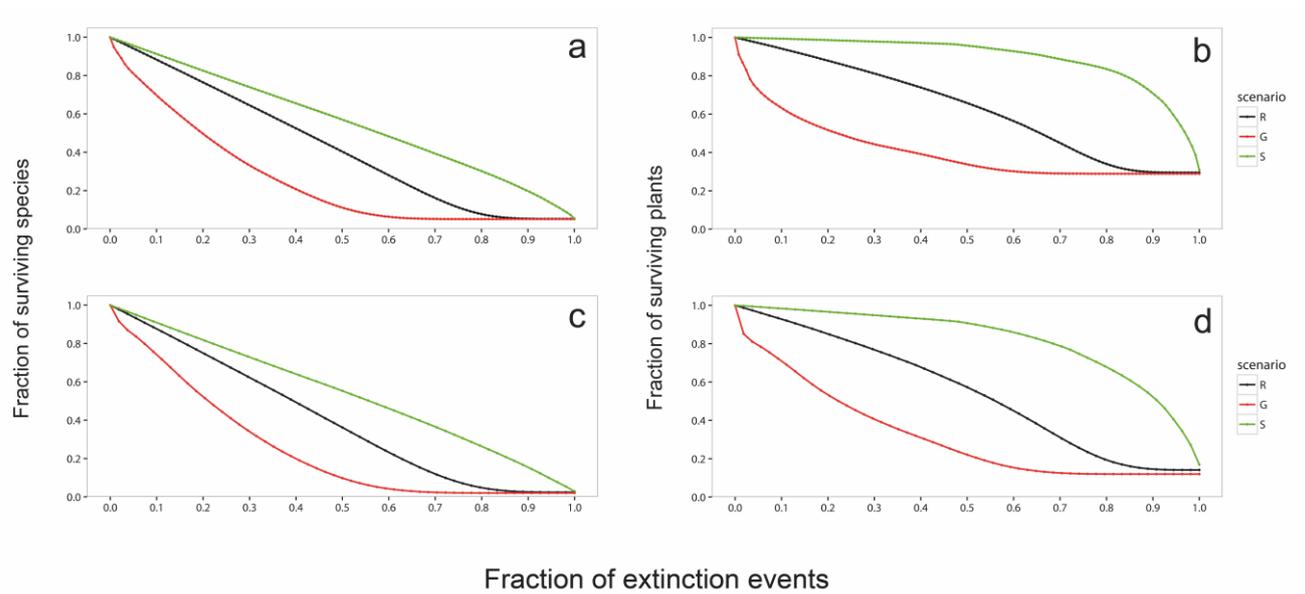